\documentstyle[twocolumn,eqsecnum,aps]{revtex}
\def\be{\begin{equation}}
\def\ee{\end{equation}}
\def\bea{\begin{eqnarray}}
\def\eea{\end{eqnarray}}
\def\eps{\varepsilon}
\begin{document}
\draft
\title{Decoupling of Zero-Modes and Covariance 
in the Light-Front Formulation of
Supersymmetric Theories}
\author{M. Burkardt}
\address{Department of Physics\\
New Mexico State University\\
Las Cruces, NM 88003-0001\\U.S.A.}
\author{F. Antonuccio}
\address{Department of Physics\\
The Ohio State University\\
Columbus, OH 43210\\U.S.A.}
\author{S. Tsujimaru}
\address{Max-Planck-Institut f\"ur Kernphysik\\
Saupfercheckweg 1\\
D-69117 Heidelberg\\
Germany}
\maketitle
\begin{abstract}
We show under suitable assumptions that
zero-modes decouple from the dynamics of non-zero
modes in the light-front formulation of some supersymmetric 
field theories. The implications for Lorentz invariance are 
discussed.
\end{abstract}
\narrowtext
\section{Introduction}
Although field theories quantized on the 
light-front (LF) have been studied for many years 
(see \cite{pb85,osu:all} and also \cite{bpp98,brazil} for a review),
recent
developments in non-perturbative string theory have 
generated additional interest. The first
surprise was M(atrix) theory \cite{bfss97}, which
was conjectured to be a non-perturbative description
of M-theory formulated in the infinite momentum frame.
Motl and Susskind provided additional insight by suggesting
that the finite $N$ version of matrix theory was 
in fact the discrete light-cone quantization (DLCQ) 
of M-theory \cite{suss97}.

Soon afterwards, the validity
of the matrix theory conjecture 
was seemingly strengthened by the 
works of Seiberg and Sen \cite{nati97,sen97},
but it was pointed out by Hellerman and Polchinski \cite{hell97} 
that a correct interpretation of their results 
required a detailed understanding of 
the (typically complicated) dynamics of zero longitudinal 
momentum modes in
the light-like compactification limit.
In general, it was observed, ``DLCQ is not a free lunch''.

The question we wish to address in this paper
is the following: ``When is the light-like limit a free lunch?''
Under a reasonable class of
assumptions, we argue that 
the zero-mode degrees of freedom in some 
supersymmetric field theories decouple, 
and so omitting
them in a DLCQ calculation 
leads to no inconsistency if the decompactification limit is 
taken prior to the light-like limit.
This observation is intriguing, since it suggests 
that the complicated zero-mode degrees of freedom studied 
in \cite{hell97} might become totally irrelevant
in the continuum limit if enough supersymmetry is present. Moreover, 
the ``correctness of matrix theory'' argument provided
by Seiberg may depend on this special property
of supersymmetric theories.

Another issue that we address is Lorentz invariance.
We show that in the light-front formulation, 
Lorentz invariance is maintained after a careful treatment
of zero modes. However, for the special case of
supersymmetric theories, the boson and fermion zero
modes that ensure Lorentz symmetry
cancel at least perturbatively! Thus, we are free
to exclude them from the outset. 

All of these observations suggest that 
the implementation of DLCQ in the absence of zero modes 
yields no inconsistency for 
supersymmetric theories.
In general, however, one needs to integrate out the zero-mode
degrees of freedom to derive an effective Hamiltonian. 
We discuss these 
issues next.

\section{Tadpole improved light-front quantization}
It has been known for a long time that field theories quantized on a
light-front $x^+ \equiv \left(x^0 + x^3\right)/\sqrt{2} =0$ leads to
a subtle  treatment of the zero modes (modes which are independent
of $x^-\equiv \left(x^0 - x^3\right)/\sqrt{2}$ \cite{my}). This result
holds both in the continuum, when zero-modes are discarded but also
in DLCQ  when the theory is 
formulated in a finite ``box'' in the $x^-$
direction with periodic boundary conditions.

Various schemes have been invented to define LF  quantization through
a limiting procedure in order to investigate these issues. For example,
one can study LF perturbation theory by starting from covariant
Feynman diagram expressions and then ``derive'' the LF Hamiltonian
perturbation theory by carefully integrating over all energies $k^-$
in loop integrations first (see for example Refs. \cite{mb:rot,mb:sg} 
and references therein).
An alternative prescription starts by quantizing the fields on a
near light-like surface (using so-called $\varepsilon$-coordinates)
and then studying the evolution of the states as one takes the LF-limit
in an infinite volume \cite{eps}.

The basic upshot of these investigations is that, at least for
theories without massless degrees of freedom, zero-modes become
high-energy degrees of freedom and ``freeze out''. However, this
does not mean that zero-modes disappear completely, since there
is still a strong interaction present among the zero-modes, giving rise
to non-trivial vacuum structure even in the LF limit.
Nevertheless, because of the high energy scale for excitations
within the zero-mode sector, one
has been able to derive {\it effective LF Hamiltonians}, where the
zero-modes have been integrated out, which act only on 
non-zero-mode degrees of freedom. Thus even though these effective
LF Hamiltonians contain only non-zero-mode degrees of freedom,
they yield the same Green's functions as a covariant calculation
provided one considers only Green's functions where all external
momenta have $k^+\neq 0$.

\subsection{Self-interacting scalar fields}
\begin{figure}
\unitlength1.cm
\begin{picture}(5,5.3)(0,-7.5)
\includegraphics{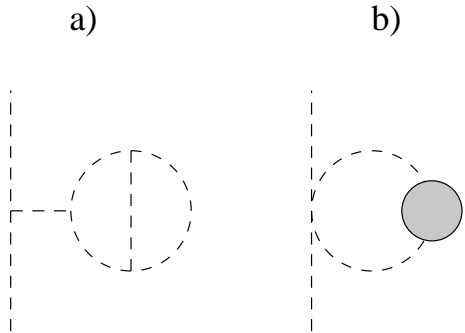}
\end{picture}
\caption{Typical self-energy diagrams for scalar fields which
vanish if zero-modes are not included. a) ``tennis racket'' shaped
tadpole in $\phi^3$ theory, b) generalized tadpole diagram in $\phi^4$
theory. The grey blob represents an arbitrary self-energy insertion.}
\label{fig:tennis}
\end{figure}

As an example, let us consider a scalar field theory with cubic
(plus higher order) self-interactions. The presence of cubic
self interactions gives rise to ``tennis racket'' Feynman diagrams 
(Fig. \ref{fig:tennis}a). 
If zero-modes are excluded then obviously
all tennis racket diagrams (which do contribute to Feynman perturbation
theory) have no analog in LF Hamiltonian perturbation theory.
However, the crucial observation is that tennis racket diagrams
are momentum independent and only lead to a mass renormalization
proportional to $\langle 0 |\phi |0\rangle$. Similarly, all
tennis racket insertions into $n$-point interactions
only lead to a
renormalization of the $(n-1)$-point interaction term, i.e. all
these diagrams can be easily integrated out.

More generally, one can show that for self-interacting scalar fields,
\footnote{Here and in the following we will implicitly restrict
ourselves
to Green's functions where all external momenta have a non-vanishing
plus-component.} zero-modes contribute only to diagrams with generalized
tadpole topology. As a result, zero-modes can be integrated out 
easily. For a polynomial self-interaction
\begin{equation}
{\cal L} = \frac{1}{2}\partial_\mu \phi \partial^\mu \phi - V(\phi)
\label{eq:lager}
\end{equation}
where
\be
V(\phi) = \sum_{k\leq n} c_k \frac{\phi^k}{k!}
\ee
one thus finds for the effective LF Hamiltonian \cite{mb:sg}
\be
P^- = V_{eff}(\phi),
\label{eq:Peff}
\ee
where the effective potential 
is also a polynomial of the same degree
\be
V_{eff}(\phi)=\sum_{k\leq n} c_k^{eff} \frac{\phi^k}{k!} 
\label{eq:Veff}
\ee
but with coefficients that are renormalized due to integrating out
zero-mode degrees of freedom
\be
c_k^{eff} = \sum_{l=k}^n c_l \langle
0|\frac{\phi^{l-k}}{(l-k)!}|0\rangle .
\label{eq:ceff}
\ee
As an illustration of how Eq. (\ref{eq:ceff}) arises, let us consider
a theory with quartic self-interactions, i.e.
$V(\phi )= \frac{1}{2} \mu^2 \phi^2 + \frac{1}{4!}\lambda^2 \phi^4$.
\footnote{For the general case, see Ref. \cite{mb:sg}.}
In this case, the only Feynman diagrams which are improperly
handled (they are set to zero!)
when the zero-mode region ($k^+=0$) is cut out are the
generalized tadpole diagrams (Fig. \ref{fig:tennis}b).
In order to see why these diagrams give only a zero-mode contribution,
let us consider the sum of all generalized tadpole diagrams, which
can be easily done by using the full propagator for the scalar fields
for which we write down a spectral representation \cite{bd}
\be
\Delta_F(p) = \int_0^\infty \! dM^2\, \frac{i \rho(M^2)}
{p^2-M^2+i\eps}
\label{eq:bspectr}
\ee
with spectral density $\rho (M^2)$.

As a side remark, for later use, we would like to point out that the 
spectral density has a very simple
representation in terms of the LF Fock states.
Upon inserting a complete set of eigenstates of the LF 
Hamiltonian into the scalar two-point function \cite{bd}, one finds
(Appendix A)
\bea
\rho (M^2) &=& 2\pi \sum_n \delta\left(\frac{M^2}{2P^+} - P_n^-\right)
\left|\langle 0|\phi (0)|n,P^+\rangle 
\right|^2
\nonumber\\
&=&2\pi \sum_n \delta\left(M^2 - M_n^2\right)
2P^+ \left|\langle 0|\phi (0)|n,P^+\rangle 
\right|^2
\nonumber\\
&=& \sum_n 
\delta\left(M^2 - M_n^2\right) b_n
\label{eq:bspectr2}
\eea
where $|n,P^+ \rangle$ is a complete set of eigenstates of $P^-$
(with eigenvalues $P_n^-=\frac{M_n^2}{2P^+_n}$) 
which we take to be normalized to $1$ and
where $b_n$ is the probability that the state $n$ is in its one boson
Fock component (one boson which carries the whole momentum $P^+$).
The sum can be evaluated at arbitrary but fixed total momentum
$P^+$ (assuming we work in the continuum limit).

Using Eq. (\ref{eq:bspectr}),
one finds for the sum of all generalized tadpole diagrams
\footnote{Note that this result holds regardless whether or not
fermions pairs contribute to the spectral density of the bosons!}
\be
-i\Sigma^{tadpole} = \frac{\lambda^2}{2}
\int_0^\infty \! dM^2 \rho(M^2)\int \frac{d^2k}{(2\pi)^2}
\frac{1}{k^2-M^2+i\eps}. 
\label{eq:tadpole}
\ee
The crucial point is that for $k^+ \neq 0$, all poles lie only on one
side
of the real $k^-$ axis and the result is thus zero
(up to a contribution from the semi-circle at infinity, which disappears
if one
subtracts the one loop result).

In order to compensate for the omission of all generalized tadpole
diagrams in naive LF quantization, we thus add a counter-term
equal to the sum of all these omitted diagrams, i.e.
a calculation that omits all explicit zero-mode degrees of freedom, but
adds
a mass counter-term $\delta \Sigma = \Sigma^{tadpole}$ will give the
same results
as a calculation that includes all zero modes explicitly.
The connection with Eq. (\ref{eq:ceff}) can now be seen by noting that
the
vacuum expectation value of $\phi^2$ is (up to a combinatoric factor)
identical to the r.h.s. of Eq. (\ref{eq:tadpole}).

In summary, one finds that (for self-interacting scalar fields)
\cite{mb:sg}
\begin{itemize}
\item zero-modes contribute to n-point functions involving
only $k^+ \neq 0$ modes only through generalized tadpole (sub-)diagrams.
By generalized tadpole diagrams we mean diagrams where a sub-diagram is 
connected to the
rest of the diagram only at one single point and hence there is no
momentum transfer through
that point.
\item n-point functions calculated with the ``tadpole improved'' 
effective LF-Hamiltonian (\ref{eq:Peff},\ref{eq:Veff},\ref{eq:ceff}) 
and without explicit zero-mode degrees of freedom
is equivalent to covariant perturbation theory generated by ${\cal L}$
(\ref{eq:lager}) to all orders in perturbation theory.
\end{itemize}
\subsection{Yukawa interactions}
As a generic example for a theory with fermions, let us now consider 
a Yukawa theory with scalar couplings
\be
{\cal L} = \bar{\psi} \left( i\gamma^\mu \partial_\mu - m_F -
g\phi\right) \psi
- \frac{1}{2}\phi \left( \Box + m_B^2\right)\phi .
\ee
If zero modes are excluded then two classes of
Feynman diagrams (to be discussed below) are treated 
improperly in the LF Hamiltonian
perturbation series.

Obviously, LF theory without zero-modes cannot generate
any tadpole (i.e. tennis racket) self energies for the fermions.
Since the above Lagrangian contains a scalar Yukawa coupling, such
diagrams are
in general non-zero. Their omission in naive LF quantization can
be easily compensated by replacing
\be
m_F\longrightarrow m_F^{eff} \equiv m_F + g \langle 0|\phi|0\rangle .
\ee

The second class of diagrams which cannot be generated by a zero-mode
free LF field theory is more subtle. As an example, let us consider 
the one loop fermion self energy \footnote{For simplicity, we will write
down
the expressions only in 1+1 dimension, but it should be emphasized that
the 
conclusions are also valid in 3+1 dimension \cite{mb:rot}.}
\bea
-i\Sigma (p) &=& g^2 \!\! \int \frac{d^2k}{(2\pi)^2}
\frac{\gamma^\mu k_\mu+m_F}{k^2 - m_F^2 + i\eps}
\frac{1}{(p-k)^2 - m_B^2+ i \eps} \nonumber\\ &=& -i\Sigma_{LF} + g^2
\!\!
\int \frac{d^2k}{(2\pi)^2}
\frac{\gamma^+}{2k^+}
\frac{1}{(p-k)^2 - m_B^2+ i \eps} 
\label{eq:self}
\eea
where
\be
-i\Sigma_{LF}= g^2\!\! \int \frac{d^2k}{(2\pi)^2} 
\frac{\gamma^\mu \tilde{k}_\nu+m_F}{k^2 - m_F^2 + i\eps}
\frac{1}{(p-k)^2 - m_B^2+ i \eps} 
\label{eq:sigmaLF}
\ee
and $\tilde{k}^+=k^+$ while $\tilde{k}^- = \frac{m_F^2}{2k^+}$ is the on
mass
shell energy for the fermion. Obviously, Eq. (\ref{eq:self}) is a mere
algebraic
rewriting of the original Feynman self-energy. The important point is
that
the second term on the r.h.s. of Eq. (\ref{eq:self}) has the same pole
structure as
a tadpole diagram and thus cannot be generated by a LF Hamiltonian.
Indeed, as one can 
easily verify, second order perturbation theory with the canonical LF
Hamiltonian
yields only (the matrix elements of) $\Sigma_{LF}$ and a disagreement
between 
self-energies calculated in covariant perturbation theory and those
calculated in
LF-Hamiltonian perturbation theory (without zero-modes) emerges. Before
we proceed 
to analyze more general diagrams which suffer from a similar problem,
let us understand
intuitively how this second term arises:

In the LF formulation, not all components of the fermion field are
independent degrees
of freedom. Multiplying the Dirac equation
\be
\left( i\gamma^\mu \partial_\mu - m_F - g\phi\right) \psi =0
\ee
by $\gamma^+$ one finds that
\be
2 i \partial_- \psi_{(-)} = \left(m_F+g\phi\right) \gamma^+ \psi_{(+)},
\label{eq:constr}
\ee
where $\psi_{(\pm)} \equiv \frac{1}{2} \gamma^\mp \gamma^\pm \psi$ .
Eq. (\ref{eq:constr}) is a constraint equation and it is often used to
eliminate
the dependent component $\psi_{(-)}$ prior to quantization. This gives
rise to
``induced'' four point interactions
\be
{\cal L}^{(4)} = -g^2 \psi_{(+)}^\dagger \phi \frac{1}{i\sqrt{2}
\partial_-} \phi \psi_{(+)}
\label{eq:L4}
\ee
in the Lagrangian after eliminating the constrained field $\psi_{(-)}$
and
hence it is possible to generate ``induced tadpoles''
diagrams by contracting for example the two scalar fields in Eq.
(\ref{eq:L4}).

Before discussing the general case, it is very instructive to
investigate the one loop fermion self-energy in more detail.
First one notes that the $2^{nd}$ order perturbation theory result
(\ref{eq:sigmaLF}) is divergent at $k^+ \rightarrow 0$
\be
\Sigma_{LF} = \frac{g^2}{8\pi} \int_0^{p^+} \frac{dk^+}{k^+(p^+-k^+)}
\frac{ k^+\gamma^- + \frac{m_F^2}{2k^+}\gamma^+ + m_F}
{p^--\frac{m_F^2}{2k^+} - \frac{m_B^2}{2(p^+-k^+)} }.
\label{eq:sigmaLF2}
\ee 
This divergence is cancelled by the self-induced inertia term, which
arises from normal ordering Eq. (\ref{eq:L4})
\be
\Sigma_{n.o.} = \frac{g^2}{8\pi}\int_0^{p^+}\frac{dk^+}{k^+} ,
\ee
yielding
\bea
\Sigma_{LF}+\Sigma_{n.o.} &=& \frac{g^2}{4\pi}\int_0^1 dx
\frac{xp^\mu \gamma_\mu + m_F}{x(1-x)p^2 - m_F^2(1-x) -m_B^2 x}
\nonumber\\
&+&\frac{g^2}{4\pi} \frac{\gamma^+}{p^+} \ln \frac{m_B^2}{m_F^2}.
\label{eq:LFno}
\eea
Several important observations can be made from Eq. (\ref{eq:LFno}).
First of all, even though including the normal ordering term renders
the self-energy finite, the final result disagrees in general with
the covariantly calculated result [the first term on the r.h.s. in Eq.
(\ref{eq:LFno})]. Furthermore, the additional term breaks covariance
(parity invariance). \footnote{This fact has been used in Ref.
\cite{mb:parity} to determine the necessary counterterm non-perturbatively
by demanding covariance for physical amplitudes.}
However, most importantly, the unwanted term vanishes for $m_F=m_B$, which
indicates already a crucial cancellation between bosonic zero-modes
and fermionic zero-modes. In the rest of this paper, we will demonstrate
for the case of certain supersymmetric theories, 
that this cancellation goes beyond the one loop result.

\begin{figure}
\unitlength1.cm
\begin{picture}(5,6)(-0.5,1)
\includegraphics{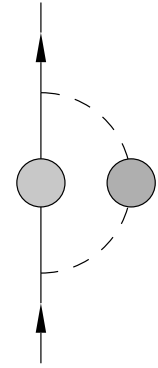}
\end{picture}
\caption{Fermion self-energy diagram, which is treated improperly
when zero-modes are excluded. The shaded blobs represent arbitrary
self-energy insertions.}
\label{fig:fself}
\end{figure}

After this more intuitive discussion of zero-mode effects for fermions,
let us now formally derive the counter-terms that arise for a theory
with Yukawa interactions.
For this purpose, it is useful to identify those Feynman diagrams
(external momenta nonzero) where zero-modes in internal lines give a
nonzero contribution to the total amplitude.
Diagrams which suffer from the same problem as the 
one-loop fermion self-energy are all diagrams where the internal lines
in
the fermion self-energy are dressed by arbitrary self-interactions
(Fig. \ref{fig:fself}).
Let us assume that all counter-terms that are necessary to
achieve agreement between LF perturbation theory (no zero modes)
and covariant perturbation theory have been added to all sub-loops
in Fig. \ref{fig:fself}, i.e. we assume that there exists a covariant
spectral representation for fermion 
propagators within the loop 
\bea
S_F(p) &=& i\int_{0}^\infty \!\!dM^2 \,\frac{\gamma^\mu p_\mu \rho_1
(M^2)
+ M \rho_2(M^2)}{p^2-M^2 + i\eps}
\label{eq:lehmann}
\eea
Similar to the scalar case, the fermion spectral density has a very
simple
representation in terms of the eigenstates of the LF Hamiltonian
as well (Appendix A)
\bea
\rho_1 (M^2) &=& \frac{2\pi}{2P^+} 
\sum_n 
\delta\left(\frac{M^2}{2P^+} - P_n^-\right) 
\left|\langle 0|\Psi_-(0)|n,P^+\rangle \right|^2
\nonumber\\
&=& \sum_n 
\delta\left(M^2 - M_n^2\right) f_n.
\label{eq:fspectr}
\eea
The spectral representation for bosons (\ref{eq:bspectr}) from the
previous section is also still valid (of course with a different
spectral function since we 
now deal with a different theory).
For later use, we also note that completeness of the LF eigenstates 
implies the normalization condition
\be
\int_0^\infty dM^2 \rho_1(M^2) = \int_0^\infty d\mu^2 \rho (\mu^2) =1
\label{eq:norm}
\ee
for the spectral densities. 

Using the above spectral representations 
[Eqs. (\ref{eq:bspectr}) and (\ref{eq:fspectr})] for the internal
propagators, we now calculate the necessary counter-term
self consistently. The covariant self-energy for the diagram in Fig.
\ref{fig:fself}
thus reads
\bea
-i\Sigma_F &=& g^2 \!\!\int_0^\infty \!\!dM^2 \int_0^\infty \!\!d\mu^2
\int 
\frac{d^2k}{(2\pi)^2}
\frac{\gamma^\nu k_\nu \rho_1(M^2)+M\rho_2(M^2)}{k^2 - M^2 + i\eps}
\nonumber\\
& &\quad \times
\frac{\rho(\mu^2)}{(p-k)^2 - \mu^2+ i \eps} .
\label{eq:sigmacov}
\eea
We will now calculate the piece which is missed when the vicinity
of both $k^+=0$ and $p^+-k^+=0$ is omitted in the integration 
in Eq. (\ref{eq:sigmacov})
(naive LF quantization with omission of fermion and boson zero-modes
respectively).
Using the one-loop analysis as a guide, it is clear that the only
problems arise in the $\gamma^+$ component of the self-energy.
In order to further isolate the troublemaker,
we use the algebraic identity 
\bea
& &\!\!\!\!\frac{k^-}{k^2-M^2+i\eps} \frac{1}{(p-k)^2 - \mu^2 +i\eps} 
=
\nonumber\\
& &\quad \quad \quad \quad
\frac{1}{2p^+} \frac{ 2(p^+-k^+)p^- + M^2-\mu^2}{
\left(k^2-M^2+i\eps \right)
\left((p-k)^2-\mu^2+i\eps \right)} \nonumber\\
& & \quad \quad \quad \quad + \frac{1}{2p^+}\left[
\frac{1}{k^2-M^2+i\eps} - \frac{1}{(p-k)^2 - \mu^2 +i\eps} \right]
.
\label{eq:alg}
\eea
Obviously, the first term on th r.h.s. of Eq. (\ref{eq:alg}) can
be straightforwardly integrated over $k^-$ and, for this term, 
the ``zero-mode
regions'' ($k^+=0$ and $p^+-k^+=0$) can be omitted without altering the
result of the integration.
However, the two last terms on the r.h.s. of Eq. (\ref{eq:alg}) have
the pole structure of simple tadpoles and hence their only contribution
to the $k$ integration is from zero-modes of the fermions $k^+ =0$
as well as the bosons $p^+-k^+=0$.
This simple observation implies that the zero-mode counter-term
from the class of diagrams in Fig. \ref{fig:fself} reads \cite{mb:adv}
\bea
-i\delta \Sigma_F &=& \frac{g^2 \gamma^+}{2p^+}
\int \frac{d^2k}{(2\pi)^2} \int_0^\infty d \mu^2 \frac{\rho (\mu^2)}
{k^2 - \mu^2 +i\eps} \nonumber\\
&-& \frac{g^2 \gamma^+}{2p^+}
\int \frac{d^2k}{(2\pi)^2} \int_0^\infty d M^2 \frac{\rho_1 (M^2)}
{k^2 - M^2 +i\eps}
\label{eq:fself0}
\eea
where we made use of the normalization of the spectral functions
(\ref{eq:norm}).

\begin{figure}
\unitlength1.cm
\begin{picture}(5,5)(-.8,1)
\includegraphics{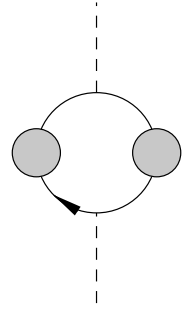}
\end{picture}
\caption{Boson self-energy diagram, which is treated improperly
when zero-modes are excluded. The full and dashed lines are fermion and
boson propagators respectively. The shaded blobs represent arbitrary
self-energy insertions.}
\label{fig:bself}
\end{figure}

A similar zero-mode counter-term arises from  ``vacuum-polarization''
type self-energies for the bosons where the fermion and anti-fermion
lines may be dressed
but where there is no interaction among the fermion and anti-fermion
(Fig. \ref{fig:bself}).
Using again the above spectral representation (\ref{eq:lehmann})
one finds for this class of Feynman diagrams
\bea
-i\Sigma &=& -2g^2 \int_0^\infty \!\!\!dM_A^2 \int_0^\infty \!\!\!dM_B^2 
\int \!\frac{d^2k}{(2\pi^2)} \label{eq:bself}\\
&\times& \frac{ k \cdot (k-p) \rho_1(M_A^2) \rho_1 (M_B^2) + M_A M_B
\rho_2 (M_A^2) \rho_2 (M_B^2)}
{\left( k^2 -M_A^2 + i\eps \right)
\left( (p-k)^2 -M_B^2 + i\eps \right)}
.\nonumber
\eea
The part of Eq. (\ref{eq:bself}) where zero-mode contributions are
crucial arise from the $k\cdot (p-k)$ term in the numerator. In order
to see this, let us write
\bea
2k\cdot (k-p) &=& \left( M_A^2 + M_B^2 - p^2 \right) \nonumber\\
& &+ \left( k^2-M_A^2 \right) +\left( (p-k)^2-M_B^2 \right) 
\label{eq:alg2} 
\eea
and we note that the first term on the r.h.s. yields no zero-mode
contribution when inserted in Eq. (\ref{eq:bself}).
However, the other two terms on the r.h.s. of Eq. (\ref{eq:alg2})
cancel one of the energy denominators and thus again yield a tadpole
like pole structure. One thus finds for the contribution from the
zero-modes
\be
-i\delta \Sigma = -g^2 \int_0^\infty dM^2 \int \frac{d^2k}{(2\pi)^2}
\frac{\rho_1 (M^2) }{k^2-M^2 + i\eps}.
\label{eq:floop0}
\ee 

Empirical studies of Feynman diagrams up to three
loops \cite{mb:rot} in Yukawa theories have shown that zero-modes play 
a role for $k^+\neq 0$ modes only in 2-point functions
(except of course through sub-diagrams).
Furthermore, of all the diagrams contributing to the two point
functions,
only the very simple sub-class of diagrams discussed above seems to
be affected when zero-modes are cut out. 
Diagrams with a more complicated topology, such as crossed diagrams
(except of course through sub-diagrams), 
require no zero-mode counter-terms when the region $k^+=0$ is cut out.
Although no rigorous analytical proof exists at this stage,
it is reasonable to assume that 
these are the only diagrams 
yielding contributions from zero-modes. 
In the following, we discuss  
the consequences of this assumption 
for supersymmetric field theories.

\section{Zero-Modes in Supersymmetric Theories}
In order to study the implications of supersymmetry on zero-mode
renormalization, let us consider a concrete example, namely a matrix
model
in 1+1 dimensions with action \cite{igor}
\be
S = \int d^2x \mbox{Tr}\left[ \frac{1}{2} \left(\partial_\mu \phi
\right)^2
+\frac{1}{2} \bar{\Psi} i \partial\!\!\!\!\!\!\not\;\, \Psi
-\frac{1}{2}V^2(\phi) -\frac{1}{2}V^\prime(\phi) \bar{\Psi}\Psi\right]
\ee
where $V(\phi) = \mu \phi - \frac{\lambda}{\sqrt{N}} \phi^2$.
The canonical LF-Hamiltonian for this model has been discussed in Ref. 
\cite{igor} and we refer the reader to this paper for details.

Obviously this model contains both three-point and four-point
interactions
for the scalar field as well as a Yukawa coupling between the scalar
field 
 and the (Majorana) fermion field, i.e. we can now directly apply above
zero-mode analysis to this model.

First we note that ``tennis racket'' tadpole diagrams (Fig.
\ref{fig:tennis}a)
must all vanish in a covariant calculation, since 
$\langle \phi \rangle \neq 0$ would break the global matrix symmetry of
the 
model. On the LF, without zero-modes, these diagrams are automatically
zero for simple kinematic reasons and
therefore, there is no need to add any zero-mode counter-terms 
for tennis racket diagrams to the LF Hamiltonian. 

Since we have already seen that all tennis racket diagrams vanish in
this 
model, the only diagrams that could still give rise to zero-mode
counter-terms
are the classes of self-energy diagrams depicted in Figs.
\ref{fig:tennis}b, \ref{fig:fself} 
and \ref{fig:bself}.

For the zero-mode contributions
to the fermion self-energy, matrix symmetry is not sufficient to prove
that
the zero-mode counter-term vanishes and we have to invoke supersymmetry.
Using the explicit expression for the supercharge $Q_-$ in terms 
of the LF fields
\be
Q_- \equiv \int dx^- :\mbox{Tr} \left[ \sqrt{2} (\partial_-\phi
  )\Psi_-\right]: 
\label{eq:Q-},
\ee
operators
we obtain the supersymmetry transformation  
\bea
&& [Q_-, \phi]=-\frac{i}{\sqrt 2}\Psi_-, \\  
&& [Q_-, \Psi_-]= \sqrt 2 \partial_{-}\phi,
\eea
which gives rise to $Q_-^2= P^+$. 
Let us show that the spectral densities $\rho$ and $\rho_1$ defined 
in the previous section is  equal 
owing to the supersymmetry. 
First note that the states of non-zero energy 
are paired by the action of supercharge. Namely, 
\bea
&&Q_-|n,P^+\rangle_{B}=\sqrt{P^+} |n,P^+\rangle_{F}, \\ 
&&Q_-|n,P^+\rangle_{F}=\sqrt{P^+} |n,P^+\rangle_{B},    
\eea
where the $B$ and $F$ denote bosonic and fermionic state, respectively. 
The fermionic state is normalized if the bosonic state 
is, i.e. $_B\langle n,P^+|n,P^+\rangle_{B}=1$ since $Q_-^2= P^+$. 
Now we can easily find 
\bea
\rho_1(M^2) &=& 2\pi  
\sum_n 
\delta\left(M^2 - M_n^2 \right) 
\left|\langle 0|\Psi_-(0)|n,P^+\rangle \right|^2
\nonumber\\
&=& 2\pi  
\sum_n 
\delta\left(M^2 - M_n^2 \right) 
\left|\langle 0|\sqrt{2} [Q_-, \phi(0)]|n,P^+\rangle \right|^2
\nonumber\\
&=& 2\pi  
\sum_n 
\delta\left(M^2 - M_n^2 \right) 2P^+
\left|\langle 0|\phi(0)|n,P^+\rangle \right|^2,  
\nonumber\\ 
&=& \rho (M^2). 
\eea

Therefore the spectral densities $\rho$ and $\rho_1$ must be equal. 

Since fermions and bosons contribute with opposite signs (but equal
strength) to the zero-mode part of the fermion self-energy 
[Eq. (\ref{eq:fself0})], the zero-mode contributions from bosons and
fermions
to the fermion self-energy cancel exactly! 
\footnote{Note that there is a flaw in the discussion of the two loop 
fermion self-energy for the SUSY Wess-Zumino model in Ref. \cite{mb:rot} 
which arises because
subtraction procedure employed in Ref. \cite{mb:rot} breaks the
supersymmetry. The unsubtracted result in Ref. 
\cite{mb:rot} is consistent with the above findings of cancellation between
bosonic and fermionic zero-mode contributions.}

The boson self-energy is more complicated, since we have to consider two
different classes of diagrams where zero-modes contribute:
tadpoles from $\phi^4$ interactions (Fig. \ref{fig:tennis}b) as well as 
the vacuum polarization type graphs (Fig. \ref{fig:bself}).
Using the results from the previous two sections, we find that the
zero-mode contribution from tadpoles to the mass reads
(\ref{eq:tadpole})
\be
\delta \mu^2 _{boson\, ZM}= 4 \lambda^2 \int_0^\infty \!\!\!dM^2
\rho(M^2) \int \frac{d^2k}
{(2\pi)^2} \frac{1}{k^2-M^2 +i\eps} .
\label{eq:bsuper}
\ee
For the contribution from zero-modes in fermion loops to the boson 
self
one finds instead
\be
\delta \mu^2 _{fermion\, ZM}= -4 \lambda^2 \int_0^\infty \!\!\!dM^2
\rho_1(M^2) \int \frac{d^2k}
{(2\pi)^2} \frac{1}{k^2-M^2 +i\eps} 
\label{eq:fsuper}
\ee
and invoking again supersymmetry, we find that the contributions from
boson
and fermion zero-modes again cancel. Note that supersymmetry has played
a dual
role in obtaining this fundamental result. First of all, it relates the 
Yukawa coupling and the scalar four-point coupling and thus the
coefficients
of Eqs. (\ref{eq:bsuper}) and (\ref{eq:fsuper}) are the same. But 
the cancellation between Eqs. (\ref{eq:bsuper}) and (\ref{eq:fsuper})
happens only because the spectral densities are the same.

\section{summary}
Even for theories with massive particles, where zero-modes are high
energy degrees of freedom, they cannot be completely discarded.
However, they can be integrated out, which gives rise to an effective
(tadpole improved) LF Hamiltonian. In supersymmetric theories, 
there is scope for 
a complete cancellation between effective interactions induced by 
bosonic zero-modes and those induced by fermionic zero-modes.
There is of course the possibility of spontaneous symmetry breaking,
in which fields acquire a non-zero expectation value. In
such a scenario, the fermion-boson cancellation may not occur, 
and we are left
with the (difficult) task of deriving an effective Hamiltonian.
However, our observations suggest that for theories with enough
supersymmetry, the zero-mode degrees of freedom may be ignored.
As a result, as long as one is interested only in the dynamics of
$k^+\neq 0$ modes in such massive supersymmetric theories, zero modes
can be discarded. This implies that for such theories DLCQ 
(in the continuum $K \rightarrow \infty$ limit, and with the zero-modes 
discarded) leads to the same Green's functions for $k^+\neq 0$ modes
as a covariant formulation. Clearly, it would be interesting
to understand the precise connection between the decoupling
of zero-modes in supersymmetric theories, and various
non-renormalizations
theorems that are known to exist. We leave this for future work.    

\acknowledgements
One of us (M.B.) was supported by the D.O.E. under contract DE-FG03-96ER40965
and in part by TJNAF.
\appendix
\section{Spectral densities}
In this appendix, we will derive some results which are useful to relate 
spectral densities to eigenstates of a LF Hamiltonian.

We start by expressing the spectral density for a scalar field 
has a simple expression in terms of the eigenstates of the LF-Hamiltonian \cite{bd}
\bea
\rho (p) = 2\pi \sum_n \int dp_n^+ \delta (p_n^+ - p^+)
\delta (p_n^- - p^-) 
\left| \langle 0 | \phi(0) | n,p^+_n \rangle \right|^2 ,
\eea
where we split up the sum over states into a sum over states at fixed momentum
$p^+_n$ and a sum (i.e. integral) over $p^+_n$.
In the next step we integrate over $p_n^+$ where we make use both of the
$\delta (p^+-p_n^+)$  as well as the relation between the LF-energy of the state and its
invariant mass $p_n^-=\frac{M_n^2}{2p_n^+}$, yielding
\bea
\rho(p) &=& 2\pi \sum_n\delta (\frac{M_n^2}{2p^+} - p^-) 
\left| \langle 0 | \phi(0) | n,p^+ \rangle \right|^2
\nonumber\\
&=&
4\pi p^+ \sum_n \delta (M_n^2 - 2p^-p^+)
\left| \langle 0 | \phi(0) | n,p^+ \rangle \right|^2 .
\label{eq:phi1}  
\eea
In order to relate Eq. (\ref{eq:phi1}) to the Fock expansion of the eigenstates
$| n,p^+ \rangle $ we use the expansion of $\phi(0)$ in terms of elementary
raising and lowering operators.
For a real scalar field, the canonical commutation relations at equal LF time
\be
\left[ \phi(x^-), \partial_- \phi (y^-) \right] = \frac{i}{2} \delta (x^--y^-)
\ee
are satisfied if one expands 
\be
\phi (x^-,x^+=0) = \int_0^\infty \frac{dk^+}{\sqrt{4\pi k^+}}
\left[ a_{k^+} e^{-ik^+x^-} + a_{k^+}^\dagger e^{ik^+x^-} \right] ,
\label{eq:phi}
\ee
where $\left[a_{k^+}, a^\dagger_{q^+}\right] = \delta (k^+-q^+)$ with all other
commutators vanishing.
Inserting Eq. (\ref{eq:phi}) into Eq. (\ref{eq:phi1}) one thus finds
\be
\rho (q)=\sum_n \delta (M_n^2 - 2p^-p^+) b_n ,
\label{eq:bn}
\ee
where 
\bea
b_n &\equiv& 4\pi p^+ \left| \langle 0 | \phi(0) | n,p^+ \rangle \right|^2
\nonumber\\
&=& \left| \langle 0 |\int_0^\infty dk^+ a_k | n,p^+ \rangle \right|^2
\eea
is the probability to find the state $| n,p^+ \rangle $
in the one boson Fock component (note that $b_n$ is $p^+$ independent!).

For the spectral density $\rho_1$ entering the full fermion propagator a similar
result can be derived. Using the representation
\be
\gamma^0 = \left( \begin{array}{cc} 0 & -i \\ i & 0 \end{array} \right)
\quad \quad \quad \quad
\gamma^1 = \left( \begin{array}{cc} 0 & i \\ i & 0 \end{array} \right)
\ee
i.e.
\be
\gamma^+ \equiv \frac{\gamma^0+\gamma^1}{\sqrt{2}}
= \sqrt{2}\left( \begin{array}{cc} 0 & 0 \\ i & 0 \end{array} \right)
\ee
one finds for the ``kinetic energy'' of a canonical Dirac field 
\be
{\cal L} = \bar{\Psi} i \gamma^+\partial_+ \Psi + ... =
\sqrt{2} \Psi_-^\dagger i \partial_+ \Psi_- + ...
\label{eq:Lkin}
\ee
where
\be
\Psi = \left( \begin{array}{c} \Psi_- \\ \Psi_+ \end{array}\right) .
\ee
Eq. (\ref{eq:Lkin}) implies
\be
\sqrt{2} \left\{ \Psi_-, \Psi_-^\dagger \right\} = \delta (x^--y^-)
\ee
and hence ($x^+=0$)
\be
\Psi_-(x^-) = 2^{-\frac{1}{4}} \int_0^\infty \frac{dk^+}{\sqrt{2\pi}} 
\left[ b_k e^{-ik^+x^-} + d_k^\dagger e^{ik^+x^-} \right] ,
\ee
where $b_{k^+}$ and $d_{k^+}$ satisfy the anti-commutation relations
\be
\left\{b_{k^+},d^\dagger_{q^+}\right\}=   
\left\{d_{k^+},d^\dagger_{q^+}\right\}=\delta (k^+-q^+) .   
\ee
In order to use this result to obtain a representation of spectral densities in
terms of the LF eigenstates, we start from the definition \cite{bd}
\bea
\left(\rho_1 p_\mu \gamma^\mu + \rho_2\right)_{\alpha \beta} &=& 2\pi
\sum_n \int dp_n^+ \delta (p_n^+ - p^+)
\delta (p_n^- - p^-) \nonumber\\
& & \quad \times\langle 0 | \Psi_\alpha (0) | n,p^+_n \rangle
\langle n,p^+_n | \bar{\Psi}_\beta (0) | 0 \rangle ,
\eea
multiply by $\gamma^+$ and take the (Dirac-) trace in order to project
out $\rho_1$, yielding
\bea 2p^+ \rho_1 (p)&=& 2\pi
\sum_n \int dp_n^+ \delta (p_n^+ - p^+)
\delta (p_n^- - p^-) \nonumber\\
& &\quad \times \sqrt{2} \langle 0 | \Psi_- (0) | n,p^+_n \rangle
\langle n,p^+_n | \Psi^\dagger_- (0) | 0 \rangle
\nonumber\\
&=& 4 \pi p^+ \sum_n \delta (M_n^2 - 2p^-p^+)
\left|\langle 0 | \Psi_- (0) | n,p^+_n \rangle \right|^2 
\eea
and therefore
\be
\rho_1 (p) =\sum_n \delta (M_n^2 - 2p^-p^+) f_n ,
\label{eq:fn}
\ee
where 
\be
f_n \equiv \left| \langle 0 | \int_0^\infty dk^+ b_k | n,p^+_n 
\rangle \right|^2
\ee
is the ($p^+$ independent!) probability for the state $n$ to be in its 
one fermion Fock component.

\end{document}